\documentclass[cameraready]{Interspeech}

\title{Integrating Facial Generation into Full-Duplex Spoken Dialogue Systems}

\author{Jingjing}{Jiang}
\author{Atsumoto}{Ohashi}
\author{Ryuichiro}{Higashinaka}

\address{Graduate School of Informatics, Nagoya University, Japan}

\email{jiang.jingjing.k6@s.mail.nagoya-u.ac.jp\\
ohashi.atsumoto.c0@nagoya-u.jp\\
higashinaka@i.nagoya-u.ac.jp}


\usepackage{comment}
\usepackage[table]{xcolor}
\newcommand{\customParagraph}[1]{{\noindent \textbf{#1. }}} 
\usepackage{booktabs}
\usepackage{arydshln}
\usepackage{amssymb}
\usepackage{amsmath}
\usepackage{graphics}
\newcommand{\newcheckmark}{\raisebox{0.6ex}{\scalebox{0.7}{$\sqrt{}$}}}

\usepackage{xcolor}
\usepackage{soul}
\setuldepth{a}

\begin{document}
\maketitle

\begin{abstract}
Full-duplex spoken dialogue models, such as Moshi, enable natural, low-latency voice conversations. However, they remain limited to the audio modality, lacking the facial expressions that are integral to human communication. We present Moshi-Face, the first full-duplex dialogue model that jointly processes the user's audio and facial input while simultaneously generating speech and facial motion. We first construct a vector-quantized variational autoencoder (VQ-VAE) as a face codec that encodes 3D head meshes extracted from facial videos into compact discrete tokens, referred to as face tokens, and conversely reconstructs 3D meshes from these tokens. We then extend Moshi with a Face Transformer module that generates face tokens non-autoregressively, enabling Moshi-Face to produce synchronized audio and face tokens in real time. Experiments show that Moshi-Face achieves audiovisual alignment at low latency while preserving the dialogue quality of the original audio-only model.
\end{abstract}

\section{Introduction}
Human conversation is inherently simultaneous and multimodal. It unfolds as a real-time, bidirectional exchange marked by overlapping, interruptions, and backchannels, and it is conveyed through both verbal and nonverbal cues, such as facial expressions, head movements, and gestures~\cite{hall2019nonverbal}. Developing dialogue systems that capture both characteristics is therefore essential for achieving natural human-computer interaction.

Several multimodal spoken dialogue systems capable of understanding and generating facial behavior have been proposed~\cite{park2024let, hu2025unitalker, see2025tan}, demonstrating that incorporating facial modalities enhances engagement, perceived naturalness, and communicative clarity in human-computer interaction~\cite{hu2025unitalker, see2025tan}. However, these systems are built upon turn-based dialogue architectures, in which only one party may speak at a time~\cite{arora2025landscape,ji2024wavchat}. This constraint prevents natural conversational dynamics such as simultaneous speech and real-time backchanneling, resulting in interactions that differ fundamentally from how humans converse.

Compared with turn-based systems, recently emerging full-duplex spoken dialogue models represent a significant advance in voice interaction~\cite{defossez2024moshi, wang2025freeze, nguyen2023generative, ohashi2025towards}. By processing two audio streams, one from the system and one from the user, in real time, these models capture the simultaneous, bidirectional dynamics characteristic of natural human conversation. Nevertheless, existing full-duplex systems are confined to the audio modality and lack the ability to process or generate facial behavior, including lip movements, facial expressions, and head motion, all of which are integral to face-to-face communication.

The goal of this work is to realize a full-duplex multimodal dialogue system capable of simultaneously processing and generating both speech and facial motion. To this end, we propose Moshi-Face, an extension of Moshi~\cite{defossez2024moshi}, one of the leading full-duplex spoken dialogue models, to support real-time generation of face tokens that drive 3D facial motion. Following Moshi's approach of representing speech waveforms as discrete tokens at a low frame rate, we introduce a face codec that encodes facial motion into a sequence of face tokens at the same frame rate as the audio tokens and reconstructs facial motion from these tokens. We further introduce a Face Transformer that generates face tokens non-autoregressively in real time, conditioned on the dialogue model's internal hidden states and spoken content.

\begin{figure}[t]
\centering
\includegraphics[width=0.9\linewidth]{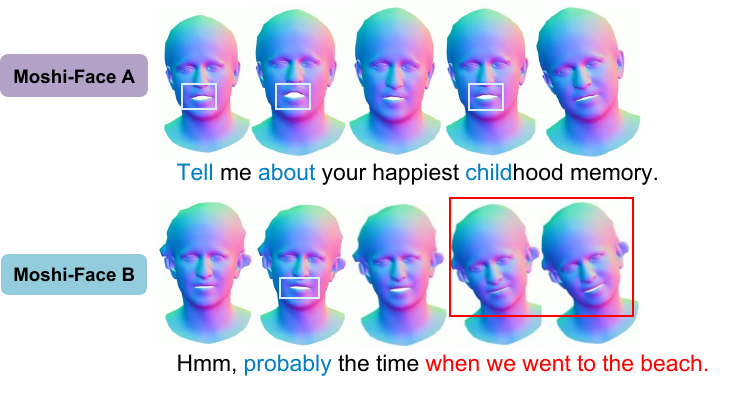}
\caption{Dialogue example generated by two Moshi-Face models. Visualized facial motion exhibits both \textcolor{cyan!60!blue}{lip synchronization} with spoken content and natural \textcolor{red}{head motion}.}
\label{fig:sample}
\end{figure}

Through experiments on 180 hours of dialogue data derived from the Seamless Interaction dataset~\cite{agrawal2025seamless}, with 3D facial motion extracted using VHAP~\cite{qian2024gaussianavatars}, we demonstrate that Moshi-Face achieves accurate audiovisual synchronization at low latency while preserving the dialogue quality of the original audio-only model. Figure~\ref{fig:sample} shows an example dialogue generated interactively by two Moshi-Face models, where the facial motion exhibits both lip synchronization with the spoken content and natural head motion.

\begin{figure*}[!t]
\centering
\includegraphics[width=1\linewidth]{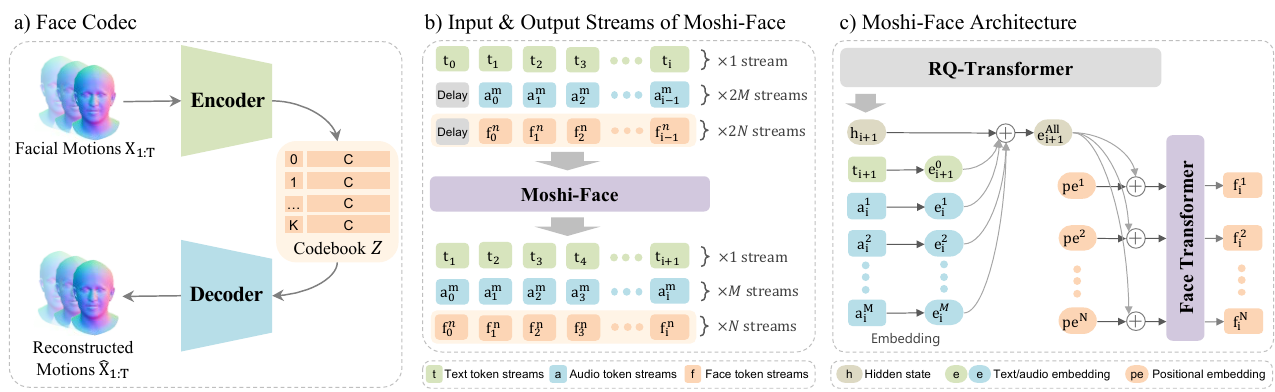}
\caption{\textbf{Overview of Moshi-Face.} a)~Face codec encodes facial motion into $N$ discrete face tokens and decodes them back. b)~Moshi-Face appends $N$ face token streams to existing text and audio token streams. c)~Face Transformer generates $N$ face tokens from conditioning vector $\mathbf{e}^{\mathrm{All}}_{i+1}$ that aggregates hidden state, text, and audio embeddings.}
\label{fig:overall}
\end{figure*}

\section{Moshi-Face}
Our goal is to extend a full-duplex spoken dialogue model to jointly process and generate 3D facial motion alongside speech. We propose \textbf{Moshi-Face}, an extension of Moshi~\cite{defossez2024moshi} that represents facial motions as discrete tokens and handles them jointly with audio in real time. As illustrated in Figure~\ref{fig:overall}, this section describes: a)~a VQ-VAE serving as a codec for 3D facial motion, b)~the input and output streams of Moshi-Face, and c)~the architecture of Moshi-Face.

\subsection{Face Codec}
\label{sec:vqvae}
To obtain discrete face tokens for the downstream Moshi-Face, we adopt a VQ-VAE~\cite{vqvae2017neural} architecture as our face codec, following the design of CodeTalker~\cite{xing2023codetalker}. As shown in Figure~\ref{fig:overall}~(a), the face codec consists of an encoder $E$, a quantizer $Q$ with a codebook $\mathcal{Z}$, and a decoder $D$. We denote the input sequence of facial motions as $\mathbf{X}_{1:T} \in \mathbb{R}^{T \times V \times 3}$, where $T$ is the number of frames, $V$ is the number of FLAME facial vertices, and $3$ corresponds to the spatial coordinates ($x, y, z$)~\cite{FLAME2017SiggraphAsia}. Each frame represents the displacement between the current vertices and static neutral facial expression.

The encoder $E$, composed of a downsampling 1-D convolutional layer followed by a transformer layer, temporally downsamples the input by a factor of $r = T/T'$ and produces $N$ latent vectors per frame, yielding $\mathbf{Z} = E(\mathbf{X}_{1:T}) \in \mathbb{R}^{T'N \times C}$, i.e., $N$ vectors for each of the $T'$ downsampled frames, where $C$ is the embedding dimension. The quantizer $Q$ maps each latent vector independently to its nearest entry in the codebook $\mathcal{Z} \in \mathbb{R}^{K \times C}$, where $K$ is the codebook size. This quantization yields $N$ discrete indices per frame, which we refer to as \emph{face tokens}; we set $N{=}8$ in our experiments. The decoder $D$, mirroring the encoder, takes the quantized sequence $\mathbf{Z}_q \in \mathbb{R}^{T'N \times C}$ and upsamples it to reconstruct the motion $\hat{\mathbf{X}}_{1:T} = D(\mathbf{Z}_q)$.

The VQ-VAE is trained with a combination of an L1 reconstruction loss, a quantization loss, and a velocity loss: $\mathcal{L}_{\mathrm{vq}} = \mathcal{L}_{\mathrm{rec}} + \lambda_{\mathrm{q}} \mathcal{L}_{\mathrm{q}} + \lambda_{\mathrm{vel}} \mathcal{L}_{\mathrm{vel}}$, where $\lambda_{\text{q}}$ and $\lambda_{\text{vel}}$ are the loss weights for the quantization loss and velocity loss, respectively.

\subsection{Input and Output Streams of Moshi-Face}
Moshi-Face appends face token streams to Moshi's existing text and audio token streams (Figure~\ref{fig:overall}~(b)). At each timestep $i$, the model operates on three types of token streams at 12.5\,Hz, taking as input the tokens of both the system and user and generating only those of the system:

\begin{itemize}
\item \textbf{Text token stream} ($1$ in, $1$ out): The text token $t_i$ represents the textual content to be spoken as the system's inner monologue.
\item \textbf{Audio token streams} ($2M$ in, $M$ out): Each speaker is represented by $M{=}8$ audio tokens processed in parallel, so the model reads $2M$ streams (system and user) and generates the $M$ streams of the system's speech. These tokens are encoded from raw waveforms at 12.5\,Hz by the encoder of Mimi, a VQ-VAE-based neural speech codec~\cite{sound2022,Neural2023}, and decoded back into waveforms by its decoder.
\item \textbf{Face token streams} ($2N$ in, $N$ out): Analogously, each speaker is represented by $N{=}8$ face tokens. The model reads $2N$ streams (system and user) and generates the $N$ streams of the system's facial motion. These tokens are encoded from 3D facial motion by the encoder of the Face codec and decoded back into motion by its decoder. 
\end{itemize}
Consequently, we train face tokens at the same frame rate as the text and audio tokens, enabling real-time input and generation of face tokens. Following the implementation of Moshi, the audio and face token streams are delayed by one timestep relative to the text tokens to improve the generation stability.

\subsection{Moshi-Face Architecture}
Moshi-Face consists of two primary components: the Residual Quantized (RQ)-Transformer and the Face Transformer (Figure~\ref{fig:overall}~(c)).

\customParagraph{RQ-Transformer} The RQ-Transformer is an autoregressive language model based on a 7B-parameter decoder-only Transformer and a smaller Depth Transformer. It autoregressively generates a hidden state and text and audio tokens, conditioned on previous tokens, along the time axis:
\begin{equation}
\mathbf{h}_{i+1} ,\, \mathbf{t}_{i+1} ,\, \mathbf{a}^{1:M}_{i} = \mathcal{T}_{\mathrm{RQ}}\big(\mathbf{t}_{(\leq i)}, \, \mathbf{a}^{1:M}_{(< i)}, \, \mathbf{f}^{1:N}_{(< i)} \big),
\label{eq:rq}
\end{equation}
where $\mathbf{t}_{(\leq i)},\, \mathbf{a}^{1:M}_{(< i)},\, \mathbf{f}^{1:N}_{(< i)}$ denote the text, audio, and face tokens. The resulting hidden state $\mathbf{h}_{i+1}$ and tokens $\mathbf{t}_{i+1},\, \mathbf{a}^{1:M}_{i}$ then condition the Face Transformer.

\customParagraph{Face Transformer}
\label{sec:face-transformer}
This module is a non-causal, transformer-based module that generates $N$ face tokens at each timestep, conditioned on the outputs of RQ-Transformer. Since the $N$ face tokens within each frame are quantized independently with no sequential dependency (Section~\ref{sec:vqvae}), they are generated non-autoregressively in parallel. At each timestep, the module first forms a unified conditioning vector by summing the hidden state with the text and audio token embeddings:
\begin{equation}
    \mathbf{e}^{\mathrm{All}}_{i+1} = \mathbf{h}_{i+1} + \mathbf{e}^0_{i+1} + \sum_{m=1}^M \mathbf{e}_i^m,
\end{equation}
where $\mathbf{e}^{0}_{i+1}$ is the text token embedding, and $\mathbf{e}^{m}_{i}$ is the embedding of the $m$-th audio token, both produced by the embedding tables of the Face Transformer. This vector is then projected and combined with a learnable positional embedding $\mathbf{pe}^n$ to form a query $\mathbf{q}^n = \mathrm{Proj}(\mathbf{e}^{\mathrm{All}}_{i+1}) + \mathbf{pe}^n$ for each of the $N$ output positions. These query vectors are then passed through the Face Transformer, which applies non-causal self-attention across the $N$ positions within each timestep, and $N$ per-component linear heads predict the face tokens $\mathbf{f}^{1:N}_i$ in parallel:
\begin{equation}
  \mathbf{f}^{1:N}_{i} = \mathcal{T}_{\mathrm{Face}}\big(\{\mathbf{q}^n\}_{n=1}^{N}\big).
  \label{eq:face}
\end{equation}
This allows all face tokens to attend to one another without any imposed ordering, consistent with the independence of the $N$ codebook indices in the face codec (Section~\ref{sec:vqvae}).

\customParagraph{Training Loss} The face generation loss $\mathcal{L}_{\mathrm{face}}$ is the cross-entropy between the predicted and ground-truth face tokens. The total training objective combines the original Moshi losses with the face generation loss: $\mathcal{L}_{\mathrm{Moshi\text{-}Face}} = \mathcal{L}_{\mathrm{text}} + \mathcal{L}_{\mathrm{audio}} + \lambda \, \mathcal{L}_{\mathrm{face}}$, where $\lambda$ weights the face generation loss.

\section{Experiments}
\subsection{3D Audiovisual Dialogue Dataset}
\label{sec:dataset}
Training Moshi-Face requires a large-scale conversational audiovisual dataset with 3D face vertices. Existing widely used 3D audiovisual datasets, such as VOCASET~\cite{voca2019capture} and BIWI~\cite{biwi2010}, are non-conversational and contain only minutes of utterance recordings, making them insufficient for training audiovisual dialogue models. We therefore leveraged a subset of Meta's Seamless Interaction Dataset~\cite{agrawal2025seamless}, a large-scale collection of face-to-face multimodal dialogues. Specifically, we randomly downloaded approximately 180 hours of dialogue data, totaling around 3,400 dialogues. Each dialogue contains time-aligned speech transcriptions, separate-channel audio, and single-speaker videos for both speakers. We tokenized the text and audio using the tokenizers provided by Moshi~\cite{defossez2024moshi}. To obtain a reliable 3D facial ground truth, we used VHAP~\cite{qian2024gaussianavatars}, a high-precision pipeline for extracting 3D facial meshes from monocular video. Facial motion was extracted at 25\,fps, with each frame represented as a 3D mesh of 5,143 vertices, each with ($x, y, z$) coordinates, based on the FLAME~\cite{FLAME2017SiggraphAsia} topology.

\vspace{-0.3em}
\subsection{Training Details}
Training proceeded in two stages. In the first stage, we used 70 hours of the extracted 3D mesh data to train the face codec (Figure~\ref{fig:overall}~(a)), split into train/valid/test sets at a ratio of $8{:}1{:}1$. The trained face codec was then applied to the full 180 hours to obtain face tokens for all dialogues. In the second stage, we trained Moshi-Face (Figure~\ref{fig:overall}~(b,\,c)) on the tokenized text, audio, and face tokens. We reserved 100 dialogues that were unseen during face codec training as the test set and split the remaining 3,300 dialogues into train/valid sets at a ratio of $9{:}1$.

\customParagraph{Face Codec} The number of face components was set to $N{=}8$, and the codebook size was $K{=}256$ with an embedding dimension of $C{=}128$. To align the face tokens with Moshi's text and audio tokens encoded at 12.5\,Hz, we set the temporal downsampling factor to $r{=}2$, such that the face codec encoded the input facial motion at 25\,fps into face tokens at 12.5\,Hz. The model was trained for 70 epochs using the AdamW~\cite{adamW2018decoupled} optimizer with a learning rate of $1e{-}4$ and a batch size of 4.

\customParagraph{Moshi-Face}
\label{sec:train-moshi_face}
We initialized the RQ-Transformer with the pre-trained Moshi checkpoint~\footnote{\scriptsize{\url{https://huggingface.co/kyutai/moshiko-pytorch-bf16}}} and attached a randomly initialized Face Transformer. Training followed a two-step strategy:
\begin{itemize}
\item \textbf{Step 1:}~The RQ-Transformer was frozen, and only the Face Transformer was trained; 
\item \textbf{Step 2:}~All components were jointly fine-tuned.
\end{itemize}
Step~1 used a learning rate of $5e{-}4$, 500 training steps, and a batch size of 32. Step~2 used learning rates of $2e {-}6$/$4e{-}6$/$1e{-}5$ for the Temporal/Depth/Face Transformers, with the face loss weight $\lambda{=}1$, 1,200 training steps, and a batch size of 16. In both steps, teacher forcing was applied: ground-truth face tokens from timestep $i{-}1$ were embedded and added to the corresponding query vectors at timestep $i$, providing autoregressive conditioning across time while maintaining non-autoregressive generation across the $N$ face tokens within each timestep.
\begin{table}[t]
\caption{\textbf{Effect of codebook size and embedding dimension on face codec codebook performance.} Perplexity was normalized by codebook size. Mean Vertex Error (MVE) and Lip Vertex Error (LVE) are in units of $\times 10^{-3}$. Best results in \textbf{bold}.}
  \label{tab:vqvae}
  \centering
  \small
  \renewcommand{\arraystretch}{0.75}
  \begin{tabular}{cc ccc}
    \toprule
    \textbf{Size} & \textbf{Dim} & \textbf{Perplexity $\uparrow$} & \textbf{MVE $\downarrow$} & \textbf{LVE $\downarrow$} \\
    \midrule
    128 & 64  & 0.66 & 11.20 & 12.79 \\
    128 & 128 & \textbf{0.67} & 11.79 & 13.95 \\
    256 & 64  & 0.57 & \textbf{9.85}  & 12.40 \\
    256 & 128 & 0.66 & 9.90  & \textbf{11.77} \\
    \bottomrule
  \end{tabular}
\end{table}
\setlength{\textfloatsep}{13pt} 

\begin{table*}[!tp]
\centering
\small
\setlength{\tabcolsep}{3.5pt}
\renewcommand{\arraystretch}{0.9} 
\caption{\textbf{Comparison of audiovisual synchronization and dialogue quality.} Performance was evaluated under \textit{teacher-forced} (ground-truth conditioned) and \textit{free-run} settings. ``Streaming'': real-time streaming capability. AV sync: Audiovisual synchronization measured by LSE-D, LSE-C. Speech: speech naturalness by UTMOS. LLMAJ: semantic quality by LLM-as-a-Judge (1--5 scale), including coherence (Coh.), naturalness (Nat.), relevance (Rel.), and overall quality (Ove.). Best in \textbf{bold}, best Moshi-Face variant \ul{underlined}.}
\label{tab:main}
  \begin{tabular}{l c cc cc ccccc}
    \toprule
    & & \multicolumn{2}{c}{\textbf{Teacher-forced}} 
      & \multicolumn{6}{c}{\textbf{Free-run}} \\
    \cmidrule(lr){3-4} \cmidrule(lr){5-11}
    & & \multicolumn{2}{c}{AV Sync} 
      & \multicolumn{2}{c}{AV Sync} & Speech 
      & \multicolumn{4}{c}{LLMAJ (1--5)} \\
    \cmidrule(lr){3-4} \cmidrule(lr){5-6} \cmidrule(lr){7-8} \cmidrule(lr){8-11}
    \textbf{Model} & \textbf{Streaming} 
      & LSE-D$\downarrow$ & LSE-C$\uparrow$
      & LSE-D$\downarrow$ & LSE-C$\uparrow$ 
      & UTMOS$\uparrow$ 
      & Coh. & Nat. & Rel. & Ove. \\
    \midrule
    Moshi                  & $\newcheckmark$ & \multicolumn{2}{c}{--} 
                           & \multicolumn{2}{c}{--} & \textbf{3.08} & 3.76 & 3.73 & \textbf{4.26} & \textbf{3.85} \\
    Moshi-ft               & $\newcheckmark$ & \multicolumn{2}{c}{--} 
                           & \multicolumn{2}{c}{--} & 1.69 & 3.59 & 4.28 & 3.95 & 3.55 \\
    \midrule
    Reconstructed face & $\times$        &  \textbf{8.53} & 0.12
                           & \multicolumn{2}{c}{--} &  \multicolumn{4}{c}{--} \\
    Random face & $\times$        &  11.7 & 0.13
                           & 11.8 & 0.11 
                           &  \multicolumn{4}{c}{--} \\
    \midrule
    \textbf{Moshi-Face (Ours)} 
                           & $\newcheckmark$ 
                           & \ul{8.76} & 0.14 
                           & 11.0 & 0.16 
                           & 1.75
                           & \ul{\textbf{3.79}} & 4.52
                           & 4.24 & \ul{3.76} \\
    \quad w/o Face Transformer pre-training   
                           & $\newcheckmark$ 
                           & 9.53 & 0.13 & 10.4 & 0.14 
                           & 1.71 & 3.78 & \ul{\textbf{4.53}} & \ul{4.25} & \ul{3.76} \\
    \quad w/o full fine-tuning          
                           & $\newcheckmark$ 
                           & 11.8 & \textbf{\ul{0.16}} & 11.1 & \ul{\textbf{0.20}} 
                           & \ul{2.42} & 3.24 & 3.94 & 3.89 & 3.23 \\
    \quad w/o $t{-}1$ face token input  
                           & $\newcheckmark$ 
                           & 11.3 & 0.15 & \ul{\textbf{10.1}} & 0.09 
                           & 1.45 & 3.65 & 4.51 & 3.89 & 3.50 \\
    \bottomrule
  \end{tabular}
\end{table*}

\vspace{-0.5em}
\subsection{Evaluation on Face Codec}
The diversity and quality of the codebook directly determine the upper bound of Moshi-Face training in the subsequent stage. We therefore evaluated the face codec from two perspectives: reconstruction quality and codebook utilization.

\customParagraph{Metrics} For reconstruction quality, following~\cite{li2025wav2sem}, we report the \textbf{Mean Vertex Error (MVE)} and \textbf{Lip Vertex Error (LVE)}. MVE computes the average L2 distance between the ground truth and reconstructed vertex positions across all vertices and frames. LVE measures the same distance restricted to lip region vertices. Both metrics are reported in units of $\times 10^{-3}$ (lower is better). For codebook utilization, we report \textbf{Perplexity}, the exponential of the entropy of the codebook usage distribution, normalized by codebook size (higher is better, 1.0 indicating perfectly uniform usage). Low perplexity indicates codebook collapse, where only a small subset of entries is actively used, resulting in a limited vocabulary of face tokens.

\customParagraph{Results} We evaluated the face codec under different configurations of codebook size $K$ and embedding dimension $C$, as shown in Table~\ref{tab:vqvae}. Increasing the codebook size from 128 to 256 consistently improved both MVE and LVE, indicating that a larger codebook provides finer-grained facial motion representation. Comparing embedding dimensions, $C{=}128$ achieved higher perplexity than $C{=}64$ at the same codebook size, suggesting more uniform codebook utilization. We adopted the configuration $K{=}256$, $C{=}128$ for all subsequent experiments.

\vspace{-0.5em}
\subsection{Evaluation on Moshi-Face}
We evaluated Moshi-Face on audiovisual synchronization, speech naturalness, and semantic quality. As shown in Table~\ref{tab:main}, we compared against the following baselines:

\begin{itemize}
\item \textbf{Moshi:} the original pre-trained model without the face generation capability.
\item \textbf{Moshi-ft:} Moshi fine-tuned on the 180-hour Seamless Interaction dataset described in Section~\ref{sec:dataset}, without face tokens.
\item \textbf{Reconstructed face:} the first 30 seconds of 100 test dialogues, in which ground-truth audio tokens are decoded by Mimi and ground-truth face tokens are decoded by the trained face codec, serving as an upper bound for audiovisual synchronization.
\item \textbf{Random face:} identical audio tokens to Reconstructed face, but face tokens randomly sampled from the token distribution of the training data, serving as a lower bound.
\end{itemize}

\customParagraph{Experimental Settings} Performance was evaluated under two settings. In the \textbf{teacher-forced} setting, ground-truth audio tokens from 100 test dialogues were provided as input, isolating the face generation capability. In the \textbf{free-run} setting, two identical Moshi-Face models interacted as system and user, where the output of one model served as the input of the other at each timestep, simulating a real-world full-duplex conversation.

\customParagraph{Audiovisual Synchronization Results} We measured the lip-sync quality using the Lip Sync Error Distance (LSE-D, $\downarrow$) and Lip Sync Error Confidence (LSE-C, $\uparrow$) based on a pre-trained SyncNet~\cite{chung2016out}, following~\cite{prajwal2020lip}. Since no established metric directly evaluates audiovisual synchronization between audio and face tokens, we adopted an indirect evaluation protocol. Specifically, for fair comparison across all conditions, we decoded audio tokens with Mimi and face tokens with the trained face codec and then rendered the decoded 3D meshes into 2D video using the rendering pipeline~\cite{qian2024gaussianavatars}. LSE-D and LSE-C were then computed on the rendered videos and decoded audio.

As shown in Table~\ref{tab:main}, under the teacher-forced setting, Moshi-Face achieved an LSE-D of 8.76, approaching the upper bound Reconstructed face and surpassing the lower bound Random face. Under the free-run setting, it maintained comparable synchronization quality, showing that audiovisual alignment is preserved even during fully autoregressive generation. 

\customParagraph{Speech Naturalness and Semantic Quality Results} We evaluated the generated output from both acoustic and semantic perspectives. Speech naturalness was assessed using UTMOS $\uparrow$~\cite{saeki2022utmos} by predicting mean opinion scores. Linguistic and semantic quality was evaluated using the LLM-as-a-Judge (LLMAJ) framework~\cite{zheng2023judging}, following~\cite{abe2026effects}. The evaluation was conducted on ASR transcriptions obtained from Whisper-large-v3~\footnote{\scriptsize{\url{https://huggingface.co/openai/whisper-large-v3}}}. GPT-5-mini~\footnote{\scriptsize{\url{https://developers.openai.com/api/docs/models/gpt-5-mini}}} was used as the evaluator to rate coherence, naturalness, relevance, and overall dialogue quality on a 1--5 scale. Each sample was evaluated three times, and the averaged scores are reported.

Both Moshi-Face and Moshi-ft obtained lower UTMOS scores than the original Moshi, which is expected, given that fine-tuning on a smaller domain-specific dataset can degrade general speech quality. Notably, the ablation without full fine-tuning achieved a higher UTMOS than other Moshi-Face variants, as only training the Face Transformer preserves the original Moshi speech generation ability.

Despite the lower UTMOS, Moshi-Face achieved the highest coherence and the second-highest naturalness in LLMAJ among all models, with overall quality comparable to Moshi. This suggests that incorporating face tokens does not degrade dialogue quality and may even provide beneficial multimodal context for semantic generation.

\customParagraph{Ablation Study} We ablated three components of Moshi-Face. Removing Face Transformer pre-training (Step~1) degraded teacher-forced LSE-D from 8.76 to 9.53, confirming the importance of isolated pre-training for initialization. Removing full fine-tuning (Step~2) resulted in the worst LSE-D and lowest LLMAJ scores, demonstrating that joint fine-tuning is essential. Removing $t{-}1$ face token input improved free-run LSE-D but degraded LSE-C and UTMOS, suggesting a trade-off between error accumulation robustness and generation quality.

\section{Conclusion}
We presented Moshi-Face, which jointly processes and generates facial motion and speech within a full-duplex dialogue framework. To train this model, we derived a 180-hour 3D audiovisual dialogue dataset from the Seamless Interaction dataset. We trained a face codec that tokenizes facial motion into discrete face tokens and reconstructs facial motion with high fidelity. We further integrated a non-causal Face Transformer into Moshi's RQ-Transformer, enabling synchronized generation of audio and face tokens. Experiments demonstrated that Moshi-Face achieves audiovisual synchronization while preserving the dialogue quality of the original audio-only model.

Our work represents a step toward conversational agents that interact simultaneously and multimodally as humans do. In future work, we plan to replace our non-causal face codec with a causal, streaming codec to enable fully real-time visual input and output. We further intend to conduct human evaluations to assess perceptual quality toward real-world deployment.


\section{Acknowledgments}
This work was supported by JST Moonshot R\&D, Grant number JPMJMS2011. We used the computational resources of the supercomputer ``Flow'' at the Information Technology Center, Nagoya University.

\section{Generative AI Use Disclosure}
We clarify that generative AI tools were used to assist with editing and polishing the manuscript text. All scientific content, experimental design, implementation, and analysis were conducted entirely by the authors.

\bibliographystyle{IEEEtran}
\bibliography{mybib}

\begin{thebibliography}{10}
\providecommand{\url}[1]{#1}
\csname url@samestyle\endcsname
\providecommand{\newblock}{\relax}
\providecommand{\bibinfo}[2]{#2}
\providecommand{\BIBentrySTDinterwordspacing}{\spaceskip=0pt\relax}
\providecommand{\BIBentryALTinterwordstretchfactor}{4}
\providecommand{\BIBentryALTinterwordspacing}{\spaceskip=\fontdimen2\font plus
\BIBentryALTinterwordstretchfactor\fontdimen3\font minus \fontdimen4\font\relax}
\providecommand{\BIBforeignlanguage}[2]{{%
\expandafter\ifx\csname l@#1\endcsname\relax
\typeout{** WARNING: IEEEtran.bst: No hyphenation pattern has been}%
\typeout{** loaded for the language `#1'. Using the pattern for}%
\typeout{** the default language instead.}%
\else
\language=\csname l@#1\endcsname
\fi
#2}}
\providecommand{\BIBdecl}{\relax}
\BIBdecl

\bibitem{hall2019nonverbal}
J.~A. Hall, T.~G. Horgan, and N.~A. Murphy, ``Nonverbal communication,'' \emph{Annual Review of Psychology}, vol.~70, no. 2019, pp. 271--294, 2019.

\bibitem{park2024let}
S.~Park, C.~Kim, H.~Rha, M.~Kim, J.~Hong, J.~Yeo, and Y.~Ro, ``Let’s {G}o {R}eal {T}alk: Spoken {D}ialogue {M}odel for {F}ace-to-{F}ace {C}onversation,'' in \emph{Proceedings of the 62nd Annual Meeting of the Association for Computational Linguistics}, 2024, pp. 16\,334--16\,348.

\bibitem{hu2025unitalker}
Y.~Hu, R.~Liu, Y.~Ren, X.~Yin, and H.~Li, ``Uni{T}alker: Conversational {S}peech-{V}isual {S}ynthesis,'' in \emph{Proceedings of the 33rd ACM International Conference on Multimedia}, 2025, pp. 10\,248--10\,257.

\bibitem{see2025tan}
W.~Tan, J.~Lian, H.~Inaguma, P.~Tomasello, P.~Koehn, and X.~Ma, ``Seeing is {B}elieving: Emotion-{A}ware {A}udio-{V}isual {L}anguage {M}odeling for {E}xpressive {S}peech {G}eneration,'' in \emph{Findings of the Association for Computational Linguistics: EMNLP 2025}, 2025, pp. 2600--2617.

\bibitem{arora2025landscape}
S.~Arora, K.-W. Chang, C.-M. Chien, Y.~Peng, H.~Wu, Y.~Adi, E.~Dupoux, H.-Y. Lee, K.~Livescu, and S.~Watanabe, ``On the {L}andscape of {S}poken {L}anguage {M}odels: A {C}omprehensive {S}urvey,'' \emph{arXiv preprint arXiv:2504.08528}, 2025.

\bibitem{ji2024wavchat}
S.~Ji, Y.~Chen, M.~Fang, J.~Zuo, J.~Lu, H.~Wang, Z.~Jiang, L.~Zhou, S.~Liu, X.~Cheng \emph{et~al.}, ``Wav{C}hat: A {S}urvey of {S}poken {D}ialogue {M}odels,'' \emph{arXiv preprint arXiv:2411.13577}, 2024.

\bibitem{defossez2024moshi}
A.~D{\'e}fossez, L.~Mazar{\'e}, M.~Orsini, A.~Royer, P.~P{\'e}rez, H.~J{\'e}gou, E.~Grave, and N.~Zeghidour, ``Moshi: a speech-text foundation model for real-time dialogue,'' \emph{arXiv preprint arXiv:2410.00037}, 2024.

\bibitem{wang2025freeze}
X.~Wang, Y.~Li, C.~Fu, Y.~Zhang, Y.~Shen, L.~Xie, K.~Li, X.~Sun, and L.~Ma, ``Freeze-{O}mni: A {S}mart and {L}ow {L}atency {S}peech-to-speech {D}ialogue {M}odel with {F}rozen {LLM},'' in \emph{Proceedings of the 42nd {I}nternational {C}onference on {M}achine {L}earning}, 2025, pp. 63\,345--63\,354.

\bibitem{nguyen2023generative}
T.~A. Nguyen, E.~Kharitonov, J.~Copet, Y.~Adi, W.-N. Hsu, A.~Elkahky, P.~Tomasello, R.~Algayres, B.~Sagot, A.~Mohamed \emph{et~al.}, ``Generative {S}poken {D}ialogue {L}anguage {M}odeling,'' \emph{Transactions of the Association for Computational Linguistics}, vol.~11, pp. 250--266, 2023.

\bibitem{ohashi2025towards}
A.~Ohashi, S.~Iizuka, J.~Jiang, and R.~Higashinaka, ``Towards a {J}apanese {F}ull-duplex {S}poken {D}ialogue {S}ystem,'' in \emph{Proceedings of the 26th {INTERSPEECH} Conference}, 2025, pp. 1783--1787.

\bibitem{agrawal2025seamless}
V.~Agrawal, A.~Akinyemi, K.~Alvero, M.~Behrooz, J.~Buffalini, F.~M. Carlucci, J.~Chen, J.~Chen, Z.~Chen, S.~Cheng \emph{et~al.}, ``Seamless {I}nteraction: Dyadic {A}udiovisual {M}otion {M}odeling and {L}arge-scale {D}ataset,'' \emph{arXiv preprint arXiv:2506.22554}, 2025.

\bibitem{qian2024gaussianavatars}
S.~Qian, T.~Kirschstein, L.~Schoneveld, D.~Davoli, S.~Giebenhain, and M.~Nie{\ss}ner, ``Gaussian{A}vatars: Photorealistic {H}ead {A}vatars with {R}igged {3D} {G}aussians,'' in \emph{Proceedings of the IEEE/CVF Conference on Computer Vision and Pattern Recognition}, 2024, pp. 20\,299--20\,309.

\bibitem{vqvae2017neural}
A.~Van Den~Oord, O.~Vinyals \emph{et~al.}, ``Neural {D}iscrete {R}epresentation {L}earning,'' in \emph{Proceedings of the {T}hirty-first {A}nnual {C}onference on {N}eural {I}nformation {P}rocessing {S}ystems}, vol.~30, 2017.

\bibitem{xing2023codetalker}
J.~Xing, M.~Xia, Y.~Zhang, X.~Cun, J.~Wang, and T.-T. Wong, ``Code{T}alker: {S}peech-{D}riven 3{D} {F}acial {A}nimation with {D}iscrete {M}otion {P}rior,'' in \emph{Proceedings of the IEEE/CVF Conference on Computer Vision and Pattern Recognition}, 2023, pp. 12\,780--12\,790.

\bibitem{FLAME2017SiggraphAsia}
T.~Li, T.~Bolkart, M.~J. Black, H.~Li, and J.~Romero, ``Learning a model of facial shape and expression from {4D} scans,'' \emph{ACM Transactions on Graphics}, vol.~36, no.~6, pp. 194:1--194:17, 2017.

\bibitem{sound2022}
N.~Zeghidour, A.~Luebs, A.~Omran, J.~Skoglund, and M.~Tagliasacchi, ``Sound{S}tream: An {E}nd-to-{E}nd {N}eural {A}udio {C}odec,'' \emph{IEEE/ACM Transactions on Audio, Speech, and Language Processing}, vol.~30, pp. 495--507, 2022.

\bibitem{Neural2023}
R.~Kumar, P.~Seetharaman, A.~Luebs, I.~Kumar, and K.~Kumar, ``High-{F}idelity {A}udio {C}ompression with {I}mproved {RVQGAN},'' in \emph{Proceedings of the {T}hirty-seventh {A}nnual {C}onference on {N}eural {I}nformation {P}rocessing {S}ystems}, vol.~36, 2023, pp. 27\,980--27\,993.

\bibitem{voca2019capture}
D.~Cudeiro, T.~Bolkart, C.~Laidlaw, A.~Ranjan, and M.~J. Black, ``Capture, {L}earning, and {S}ynthesis of {3D} {S}peaking {S}tyles,'' in \emph{Proceedings of the IEEE/CVF Conference on Computer Vision and Pattern Recognition}, 2019, pp. 10\,101--10\,111.

\bibitem{biwi2010}
G.~Fanelli, J.~Gall, H.~Romsdorfer, T.~Weise, and L.~Van~Gool, ``A {3-D} {A}udio-{V}isual {C}orpus of {A}ffective {C}ommunication,'' \emph{IEEE Transactions on Multimedia}, vol.~12, no.~6, pp. 591--598, 2010.

\bibitem{adamW2018decoupled}
I.~Loshchilov and F.~Hutter, ``Decoupled {W}eight {D}ecay {R}egularization,'' in \emph{Proceedings of the 7th {I}nternational {C}onference on {L}earning {R}epresentations}, 2019.

\bibitem{li2025wav2sem}
H.~Li, J.~Dai, X.~Zhao, F.~Zhou, J.~Pan, and L.~Li, ``Wav2{S}em: {P}lug-and-{P}lay {A}udio {S}emantic {D}ecoupling for 3{D} {S}peech-{D}riven {F}acial {A}nimation,'' in \emph{Proceedings of the Computer Vision and Pattern Recognition Conference}, 2025, pp. 183--192.

\bibitem{chung2016out}
J.~S. Chung and A.~Zisserman, ``Out of time: automated lip sync in the wild,'' in \emph{Proceedings of the 13th Asian Conference on Computer Vision}, 2016, pp. 251--263.

\bibitem{prajwal2020lip}
K.~Prajwal, R.~Mukhopadhyay, V.~P. Namboodiri, and C.~Jawahar, ``A {L}ip {S}ync {E}xpert is {A}ll {Y}ou {N}eed for {S}peech to {L}ip {G}eneration in the {W}ild,'' in \emph{Proceedings of the 28th ACM International Conference on Multimedia}, 2020, pp. 484--492.

\bibitem{saeki2022utmos}
T.~Saeki, D.~Xin, W.~Nakata, T.~Koriyama, S.~Takamichi, and H.~Saruwatari, ``{UTMOS}: {UT}okyo-{S}arulab {S}ystem for {V}oice{MOS} {C}hallenge 2022,'' in \emph{Proceedings of the 23rd {INTERSPEECH} Conference}, 2022.

\bibitem{zheng2023judging}
L.~Zheng, W.-L. Chiang, Y.~Sheng, S.~Zhuang, Z.~Wu, Y.~Zhuang, Z.~Lin, Z.~Li, D.~Li, E.~Xing \emph{et~al.}, ``Judging {LLM}-as-a-{J}udge with {MT}-{B}ench and {C}hatbot {A}rena,'' in \emph{Proceedings of the {T}hirty-seventh {A}nnual {C}onference on {N}eural {I}nformation {P}rocessing {S}ystems}, vol.~36, 2023, pp. 46\,595--46\,623.

\bibitem{abe2026effects}
Y.~Abe, M.~Saeki, A.~Ohashi, S.~Takamichi, S.~Fujie, T.~Kobayashi, T.~Ogawa, and R.~Higashinaka, ``Effects of {D}ialogue {C}orpora {P}roperties on {F}ine-{T}uning a {M}oshi-{B}ased {S}poken {D}ialogue model,'' in \emph{Proceedings of the 16th International Workshop on Spoken Dialogue System Technology}, 2026, pp. 104--108.

\end{thebibliography}

\end{document}